\documentclass[12pt,letterpaper]{article}
\usepackage[pdftex]{graphicx,color}
\usepackage{hyperref}
\usepackage[utf8]{inputenc}
\usepackage{amsmath,amssymb}
\usepackage[dvips,letterpaper,text={6.5in,9in}]{geometry}
\usepackage{fancyhdr}
\usepackage{verbatim}

\usepackage{dcolumn}   
\usepackage{subfig}
\usepackage{float}
\usepackage{ytableau}
\newcommand\ltap{\
  \raise.3ex\hbox{$<$\kern-.75em\lower1ex\hbox{$\sim$}}\ }
\newcommand\gtap{\
  \raise.3ex\hbox{$>$\kern-.75em\lower1ex\hbox{$\sim$}}\ }

\newcommand\simge{\mathrel{%
   \rlap{\raise 0.511ex \hbox{$>$}}{\lower 0.511ex \hbox{$\sim$}}}}
\newcommand\simle{\mathrel{
   \rlap{\raise 0.511ex \hbox{$<$}}{\lower 0.511ex \hbox{$\sim$}}}}

\newcommand{\slashchar}[1]%
        {\kern .25em\raise.18ex\hbox{$/$}\kern-.60em #1}
\def\lsim{\mathrel{\raise.3ex\hbox{$<$\kern-.75em\lower1ex\hbox{$\sim$}}}}
\def\gsim{\mathrel{\raise.3ex\hbox{$>$\kern-.75em\lower1ex\hbox{$\sim$}}}}
\newcommand{\bs}{\boldsymbol}

\newcommand\CL{{\cal L}}

\newcommand\CO{{\cal O}}

\newcommand\be{\begin{equation}}
\newcommand\ee{\end{equation}}
\newcommand\bea{\begin{eqnarray}}
\newcommand\eea{\end{eqnarray}}
\newcommand\ba{\begin{array}}
\newcommand\ea{\end{array}}
\newcommand\nn{\nonumber}

\newcommand{\thalf}{\textstyle{\frac{1}{2}}}

\newcommand\gev{{\rm GeV}}
\newcommand\tev{{\rm TeV}}

\newcommand\fb{{\rm fb}}

\newcommand\ellm{\ell^-}

\newcommand\ellp{\ell^+}

\begin{document}

\title{
\vskip -15mm
\begin{flushright}
  \vskip -15mm
  {\small CERN-PH-TH-2015-175\\
    LAPTH-042/15\\
  } 
\vskip 5mm
\end{flushright}
{\Large{\bf Heavy Vector Partners of the Light Composite Higgs}}\\
} \author{ {\large Kenneth Lane$^{1,2}$\thanks{lane@bu.edu} \,\,and Lukas
    Pritchett$^{1}$\thanks{lpritch@bu.edu}}\\
  {\large $^{1}$Department of Physics, Boston University}\\
  {\large 590 Commonwealth Avenue, Boston, Massachusetts 02215}\\
{\large $^{2}$CERN Theory Division}\\
{\large CH-1211, Geneva 23, Switzerland}
} \maketitle


\begin{abstract}
  
  If the Higgs boson $H(125)$ is a composite due to new strong interactions
  at high energy, it has spin-one partners, $\rho_H$ and $a_H$, analogous to
  the $\rho$ and $a_1$ mesons of QCD. These bosons are heavy, their mass
  determined by the strong interaction scale. The strongly interacting
  particles light enough for $\rho_H$ and $a_H$ to decay to are the
  longitudinal weak bosons $V_L = W_L,\, Z_L$ and the Higgs boson $H$. These
  decay signatures are consistent with resonant diboson excesses recently
  reported near $2\,\tev$ by ATLAS and CMS. We calculate $\sigma\times
  BR(\rho_H \to VV) =$ few~$\fb$ and $\sigma\times BR(a_H \to VH) =
  0.5$--$1\,\fb$ at $\sqrt{s} = 8\,\tev$, increasing by a factor of 5--7 at
  $13\,\tev$. Other tests of the hypothesis of the strong-interaction nature
  of the diboson resonances are suggested.

  \end{abstract}


\newpage

\section*{1. Introduction}

The ATLAS and CMS Collaborations have reported 2--3$\sigma$ excesses in the
8-TeV data of high-mass diboson ($VV = WW,WZ,ZZ$)
production~\cite{Aad:2015owa,Khachatryan:2014gha,Khachatryan:2014hpa}. The
ATLAS excesses are in nonleptonic data (both $V \to \bar q q$ jets) in which
the boosted $V$-jet is called a $W$ ($Z$) if its mass $M_V$ is within
$13\,\gev$ of 82.4 (92.8)~GeV. They appear in all three invariant-mass
``pots'', $M_{WW}$, $M_{WZ}$ and $M_{ZZ}$, although there may be as much as
30\% spillover between neighboring pots. Perhaps not surprisingly, the
largest excess is in $M_{WZ}$. It is centered at $2\,\tev$, with a
3.4$\sigma$ local, 2.5$\sigma$ global significance. The ATLAS nonleptonic
$WZ$ excess has been estimated to correspond to a signal cross section times
branching ratio of $3\,\fb$.\footnote{G.~Brooijmans, D.~Morse and C.~Pollard,
  communication at Les Houches Workshop, {\em Physics at TeV Colliders},
  June~1--19, 2015.} The CMS papers report semileptonic ($V\to \ell\nu$ or
$\ellp\ellm$ plus $V\to \bar q q$) as well as nonleptonic $VV$ events. In the
purely nonleptonic sample, a boosted jet is called a $W$ or $Z$ candidate if
$70 < M_V < 100\,\gev$. A nonleptonic $V$-jet in the semileptonic sample is
considered a $W$-jet candidate if $65 < M_V < 105\,\gev$ and a $Z$-candidate
if $70 < M_V < 110\,\gev$.\footnote{This discussion does not do the
  selections of $W$ and $Z$ jets justice. The reader is urged to consult the
  ATLAS and CMS papers for a complete description of nonleptonic $W,Z$-jet
  identification.} The semileptonic data is divided into $WW$ and $ZZ$ pots.
There is a 1$\sigma$ excess in $WW$ and 2$\sigma$ in $ZZ$, both centered at
$1.8\,\tev$. CMS combined its semileptonic and nonleptonic data (which also
showed 1--2$\sigma$ excesses near $1.8\,\tev$, and still obtained a 2$\sigma$
effect at $1.8\,\tev$. ATLAS saw no similar excesses in its semileptonic
$VV$-data~\cite{Aad:2014xka, Aad:2015ufa}. Both experiments also looked for
$VH$ resonances. CMS reported a 2$\sigma$ excess near $1.8\,\tev$ in $WH \to
\ell\nu \bar bb$~\cite{CMSPAS:2014yyy}. ATLAS searched for $WH$ and $ZH$ in
semileptonic modes but saw no excess~\cite{Aad:2015yza}.

Despite the low statistics, 5--10 events, of the ATLAS and CMS excesses,
their number and proximity have inspired a number of theoretical papers
variously proposing them to be due to production of heavy weak $W'$ and $Z'$
bosons~\cite{Cheung:2015nha, Dobrescu:2015qna,Abe:2015uaa,Brehmer:2015cia},
of heavy vector bosons associated with new strong dynamics at the TeV scale
that is responsible for electroweak symmetry breaking~\cite{Thamm:2015csa,
  Franzosi:2015zra, Carmona:2015xaa}, or of a new heavy
scalar~\cite{Cacciapaglia:2015nga,Sanz:2015zha}.

If these excesses are confirmed in Run~2 data --- and that's a big if! ---
their most plausible explanation, in our opinion, is that they are the
lightest vector and, possibly, axial-vector triplet bound states of new
strong interactions responsible for the compositeness of the $125\,\gev$
Higgs boson $H$. If the Higgs is composite, it is widely believed to be built
of fermion-(anti)fermion pairs which carry weak isospin and whose other bound
states respect custodial $SU(2)$ symmetry (see, e.g.,
Refs.~\cite{Contino:2011np,Marzocca:2012zn,Bellazzini:2012tv,Lane:2014vca}).
Then there are isovector and isoscalar bosons analogous to the familiar
$\rho$, $\omega$ and $a_1$ mesons. In this paper we concentrate on the
isovectors, which we call $\rho_H$ and $a_H$ to emphasize their relation to
$H$. We shall explain that the only hadrons of the new interaction lighter
than $\rho_H$ and $a_H$ are the longitudinally-polarized weak bosons,
$V_L = W_L,\,Z_L$, and $H$ itself, which, therefore, are their decay
products. The production mechanisms of $\rho_H$ and $a_H$ are the Drell-Yan
(DY) process, induced by mixing with the photon, $W$ and $Z$, and weak vector
boson fusion (VBF). We find total production times decay rates of a few
femtobarns (fb), dominated by DY. The hallmark of the isovectors' underlying
strong dynamics are their large widths, dominated by decays involving
$V_L$. The diboson data favors $\Gamma(\rho_H) \simle 200\,\gev$, though a
somewhat greater width is still allowed. Production rates of $\rho_H$ more
than a few~fb typically imply larger widths. The mode $\rho_H \to V_L V_L$ is
completely dominant. The main two-body decay mode of $a_H$ is $V_L H$, while
the longitudinal-transverse mode, $V_L V_T$, and the on-mass-shell
$\rho_H V_L$ mode are much suppressed.  We have not estimated the nonresonant
three-body mode $a_H \to 3 V_L$.

Isovectors of composite Higgs dynamics and their interactions with Standard
Model (SM) particles, including the Higgs, have been anticipated in several
recent papers~\cite{Contino:2011np,Marzocca:2012zn,Bellazzini:2012tv,
  Lane:2014vca}. The models in Refs.~\cite{Contino:2011np,Marzocca:2012zn,
  Bellazzini:2012tv} and the particular model we use for describing isovector
couplings to SM particles are conveniently described by a hidden local
symmetry (HLS)~\cite{Bando:1987br} --- $SU(2)_L \otimes SU(2)_R$ with equal
gauge couplings, $g_L = g_R$. This parity is softly (spontaneously) broken.
The resulting vector and axial-vector bosons comprise two isotriplets, nearly
degenerate within each multiplet. Their dimension-three and~four
interactions, including those with electroweak (EW) gauge bosons respect this
parity up to corrections of order the EW gauge couplings.

In light composite Higgs models in which $H$ is a pseudo-Goldstone boson
(PGB) (see, e.g., Ref.~\cite{Bellazzini:2014yua,Panico:2015jxa} for a review)
the isovectors' expected mass is $\sim g_{\rho_H} f$, where
$g_{\rho_H} \simeq g_L = g_R$ and $f$~is the PGB decay constant, typically
$\CO(1\,\tev)$. 
In the model of Ref.~\cite{Lane:2014vca}, electroweak symmetry breaking is
driven {\em not} by technicolor, but by strong extended technicolor
interactions (ETC) at a scale of 100's of TeV. The Higgs boson in this
Nambu--Jona-Lasinio-like model~\cite{Nambu:1961tp,Nambu:1961fr} is not a PGB;
it is made light by fine-tuning the strength of the ETC interaction coupling
to be near the critical value for spontaneous electroweak symmetry
breaking. But ETC's unbroken subgroup, technicolor, is a confining
interaction and it binds technifermions into hadrons whose typical mass is
the technicolor scale $\Lambda_{TC} = \CO(1\,\tev)$. We can also use the HLS
formalism to describe the $\rho_H$, $a_H$ in this scenario and so, again,
their masses can be expressed as $g_{\rho_H} f$ where
$f \simeq \Lambda_{TC}$. From the earliest days of technicolor, the mass of
the technirho in a one-doublet model was estimated (naively) to be
$\sim 1.8\,\tev$~\cite{Dimopoulos:1979sp,Dimopoulos:1980yf}.

The interactions of the isovectors with $W,Z$ and $H$ are given in Sec.~2.
These are used to calculate the isovectors' decay rates and production cross
sections in Sec.~3. Finally, in Sec.~4 we make comments and predictions that
should test our composite-Higgs hypothesis in the first year or two of LHC
Run~2.

\section*{2. $\rho_H,\, a_H$ Couplings to Standard Model Particles}

In a light composite Higgs model the strongly-interacting bound states
lighter than $\rho_H$ are the quartet consisting of three Goldstone bosons,
$W_L^\pm$ and $Z_L$, and the scalar $H$. But is that all? If the model has
other PGBs they may be lighter than $\rho_H$. But then we would have to infer
that the $\rho_H$ production rate is rather larger than a few~fb to make up
for the smaller $VV$ branching ratio and that, we shall see in Sec.~3, is
difficult to accommodate in this sort of model. In the model of
Ref.~\cite{Lane:2014vca} the low-energy theory below $M_{\rho_H}$ {\em is}
the SM plus suppressed higher-dimension operators. Just above the electroweak
symmetry breaking transition, $W_L^\pm, Z_L,H$ are a light degenerate
quartet; just below it, they are three Goldstone bosons and a light
scalar. There are no other light hadrons of the strong interactions than
these four. They and, presumably, $\rho_H$ are lighter than $a_H$. To
minimize the contribution to the $S$-parameter~\cite{Kennedy:1988sn,
  Peskin:1990zt,Golden:1990ig, Holdom:1990tc, Altarelli:1991fk} from the
low-lying hadrons, we assume that $a_H$ and $\rho_H$ are nearly degenerate
with the same coupling strength to the electroweak currents (see, e.g.,
Ref.~\cite{Casalbuoni:1995qt,Lane:2009ct}). This greatly suppresses the
strong decay $a_H \to \rho_H V_L$.

The effective Lagrangian describing $\rho_H VV$ and $a_H VV$ couplings is
obtained from the HLS approach describing the isovectors as $SU(2)_L \otimes
SU(2)_R$ gauge bosons.  Refs.~\cite{Marzocca:2012zn, Bellazzini:2012tv} give
quite similar results for these couplings. We use ones like these that are
given in Sec.~VI of Ref.~\cite{Lane:2009ct}, adapted to the case of a single
technidoublet with no light PGBs, and with couplings chosen to cancel the
$\rho_H$ and $a_H$ contributions to $S$. They are:
\bea
\label{eq:rhoVV}
\CL(\rho_H \to VV) &=& -\frac{ig^2 g_{\rho_H}v^2}{2M^2_{\rho_H}}\rho^0_{H\mu\nu}
W^+_\mu W^-_\nu -\frac{ig^2 g_{\rho_H}v^2}{2M^2_{\rho_H}\cos\theta_W}
   \left(\rho^+_{H\mu\nu} W^-_\mu - \rho^-_{H\mu\nu} W^+_\mu\right)Z_\nu\,, \\
\label{eq:aVV}
\CL(a_H \to VV) &=& \frac{ig^2 g_{\rho_H}v^2}{2M^2_{\rho_H}}
a^0_{H\mu}\bigl(W^+_{\mu\nu} W^-_\nu - W^-_{\mu\nu} W^+_\nu\bigr)\nn\\
&-& \frac{ig^2 g_{\rho_H}v^2}{2M^2_{\rho_H} \cos\theta_W}
\bigl[a^+_{H\mu}\bigl(W^-_\nu Z_{\mu\nu} - W_{\mu\nu}^- Z_\nu\bigr) - {\rm
  h.c.}\bigr] \,.
\eea
Note the isospin symmetry of these couplings. Here, $G_{\mu\nu} =
\partial_\mu G_\nu - \partial_\nu G_\mu$, $g$ is the weak-$SU(2)$ coupling;
$g_{\rho_H}$ is the left-right symmetric HLS gauge coupling for the
isovectors. The $\rho_H$ mass in Ref.~\cite{Lane:2009ct} is nominally given
by $M_{\rho_H} = \thalf g_{\rho_H} f_{\rho_H}$, where $f_{\rho_H}$ is the HLS
decay constant (analogous to the decay constant of a PGB composite Higgs). If
we take $f_{\rho_H} = 1\,\tev \simeq 4v$, where $v=246\,\gev$ is the Higgs
vacuum expectation value, then $g_{\rho_H} = 4$ for $M_{\rho_H} = 2\,\tev$.

For highly-boosted weak bosons, as is the case here, $V^{\pm,0}_{L\mu} =
\partial_\mu \pi^{\pm,0}/M_V +\CO(M_V/E_V)$,
where $\pi$ is the pseudoscalar Goldstone boson eaten by~$V$. Then, the
$V_L V_L$ part of $V_{\mu\nu}$ is suppressed by $M_V^2/E_V^2$ and, while
$\rho_H \to V_L V_L$ is allowed, only the strongly suppressed
$a_H \to V_L V_T$ is. The same parity argument applies in reverse to the
decays $\rho_H, a_H \to V_LH$. Furthermore, for (nearly) degenerate $\rho_H$
and $a_H$, the two comprise parity-doubled triplets and, for a {\em light}
Higgs, the decay rates $\rho_H \to V_L V_L$ and $a_H \to V_L H$ are
identical.\footnote{More precisely, they are identical in the Wigner-Weyl
  mode of electroweak symmetry in which $(H, {\bs \pi})$ are a degenerate
  quartet. We thank T.~Appelquist for this simple argument for the $a_H VH$
  coupling strength.} Thus,
%
\be
\label{eq:aVH}
\CL(a_H \to V H) = gg_{\rho_H}v
\left(a^+_{H\mu}W^-_{\mu} + a^-_{H\mu}W^+_{\mu}\right) H
+ \frac{gg_{\rho_H}v}{\cos\theta_W}a^0_{H\mu}\, Z_\mu H\,.
\ee

The $a_H \rho_H V$ couplings are also taken from Ref.~\cite{Lane:2009ct}:
\bea\label{eq:arhoV}
\CL(a_H \to \rho_H V) &=& -\frac{igg^2_{\rho_H}v^2}{2\sqrt{2}M^2_{\rho_H}}
\bigl[a^0_{H\mu}\bigl(\rho^+_{H\mu\nu} W^-_\nu - \rho^-_{H\mu\nu}
    W^+_\nu\bigr) \nn\\
&+& a^+_{H\mu}\bigl(\rho^-_{H\mu\nu} Z_\nu/\cos\theta_W
    - \rho^0_{H\mu\nu} W^-_\nu\bigr) - {\rm h.c.} \bigr]\,.
\eea

Finally, the amplitudes for DY production of $\rho_H$, $a_H$ and their decay
to $VV$, $VH$ involve their mixing with $\gamma,W,Z$. (The $\rho_H$ and $a_H$
have no appreciable direct coupling to SM fermions in the composite Higgs
models considered here.) These are of $\CO(g M^2_{\rho_h}/g_{\rho_H},\,
g' M^2_{\rho_h}/g_{\rho_H})$ and depend on the electroweak quantum numbers of
their constituent fermions. We use the couplings of Ref.~\cite{Lane:2002sm},
appropriate to a single fermion doublet, for which we assume electric charges
$\pm \thalf$. The DY cross sections given in Ref.~\cite{Lane:2002sm} are
easily modified for the case at hand in which there are no other light PGBs.
They are encoded in {\sc Pythia}~6.4~\cite{Sjostrand:2006za}.

\section*{3. $\rho_H,\, a_H$ Decay Rates and Cross Sections}

The $\rho_H$ decay rates are completely dominated by the emission of a pair
of longitudinally-polarized weak bosons. The factor of $M_{\rho_H}^2$ from
the longitudinal polarization vectors is canceled by the $1/M_{\rho_H}^2$ in
Eq.~(\ref{eq:rhoVV}), giving (for $M_{\rho_H} \gg M_W$)
\be\label{eq:gamrhoVV}
\Gamma(\rho_H^0 \to W^+ W^-) \cong \Gamma(\rho_H^\pm \to W^\pm Z) \cong
 \frac{g_{\rho_H}^2 M_{\rho_H}}{48\pi}\,.
\ee

The $a_H \to VH$ decay rate from Eq.~(\ref{eq:aVH}) is
\be\label{eq:gamaVH}
\Gamma(a^0 \to ZH) \cong \Gamma(a^\pm \to W^\pm H) \cong
 \frac{g_{\rho_H}^2 M_{a_H}}{48\pi}\,.
\ee
As noted above, CMS, but not ATLAS, saw a $2\sigma$ excess in the $WH$ channel.
If this excess persists and is confirmed by ATLAS, in our model it must be
due to $a_H$.

The greatly suppressed decay rate of $a_H$ to a pair of weak bosons is 
%
\be
\label{eq:gamaVV}
\Gamma(a_H^0 \to W^+ W^-) \cong \Gamma(a_H^\pm \to W^\pm Z) \cong
\frac{g_{\rho_H}^2 M_W^2 M_{a_H}^3}{24\pi M_{\rho_H}^4}\,.
\ee
Finally, the decay rate for $a_H$ to individual $\rho_H V$ states is
\bea\label{eq:gamarhoV}
\Gamma(a_H \to \rho_H V) &=& \frac{g^2}{192\pi} \left(\frac{g_{\rho_H}
    v}{M_{\rho_H}}\right)^4 \frac{p^3}{(M_{a_H} M_{\rho_H} M_W)^2} \nn\\
&\qquad& \times \left[6 M_{\rho_H}^2(M_{a_H}^2 + M_V^2) + M_{\rho_H}^4 +
  M_{a_H}^2 p^2 - (M_{a_H}^2 - M_V^2)^2\right]\,,
\eea
where $p$ is the $V = W,Z$ momentum in the $a_H$ rest frame. An interesting
possibility would be that this quasi-two-body decay is not very limited by
phase space. The two weak bosons from $\rho_H$ would have $M_{VV} \simeq
M_{\rho_H}$ and the third $V$ would be soft and not included in the diboson
mass. A possibility like this was considered in
Ref.~\cite{Aguilar-Saavedra:2015rna}. Unfortunately, the $a_H \to \rho_H V$
decay rate is only a few~MeV in our model.

 \begin{table}[!t]
     \begin{center}{
  \begin{tabular}{|c|c|c|c|}
  \hline
 $M_{\rho_H}$ (GeV) & $\Gamma(\rho_H \to VV)$ (GeV) & $\Gamma(a_H \to VH)$  (GeV)&
 $\Gamma(a_H \to VV)$ (GeV) \\
  \hline\hline
 1800 & 178  & 184 & 0.82 \\ 
 1900 & 188  & 196 & 0.78 \\
 2000 & 198  & 208 & 0.74 \\
 \hline
 \end{tabular}}
 \caption{Principal decay rates of the isovector bosons $\rho_H$ and $a_H$ for
   $g_{\rho_H} = 3.862$ and $M_{a_H} = 1.05 M_{\rho_H}$.\label{tab:Decays}}
 \end{center}
 \end{table}

The decay rates are listed in Table~1 for $M_{\rho_H} = 1800$, 1900,
$2000\,\gev$ and $M_{a_H} = 1.05 M_{\rho_H}$; the strong coupling is fixed at
$g_{\rho_H} = 1900\,\gev/2v = 3.862$. The $\sim 200\,\gev$ width of $\rho_H$
is compatible with the existing data.
 
The main production mechanisms of the isovectors are DY and VBF. The cross
sections for the dominant modes, $\rho_H^{\pm,0} \to W^\pm Z$, $W^+W^-$ and
$a_H^{\pm,0} \to W^\pm H$, $ZH$, are listed in Table~2 for $M_{\rho_H} =
1800$--$2000\,\gev$, $M_{a_H} = 1.05 M_{\rho_H}$ and $g_{\rho_H} = 3.862$.
The DY and VBF rates for $\rho_H$ are given separately; VBF rates for $a_H$
are very small. No $K$-factor has been applied to the cross sections. The
rates reveal the following (all $BR \simeq 1$)
\begin{itemize}
\item $\sigma_{DY}(a_H) \simeq 0.5\,\sigma_{DY}(\rho_H)$.
\item $\sigma_{DY}(13\,\tev) = 5$-$7\,\sigma_{DY}(8\,\tev)$
\item $\sigma_{VBF}(a_H) \simle 0.01 \, \sigma_{VBF}(\rho_H)$.
\item $\sigma_{VBF}(\rho_H) \simeq
  \textstyle{\frac{1}{4}}\sigma_{DY}(\rho_H)$ at $\sqrt{s} = 8\,\tev$, rising
  to about $\textstyle{\frac{1}{2}}\sigma_{DY}(\rho_H)$ at $13\,\tev$.
\item $\sigma(\rho_H^\pm) \simeq 2\sigma(\rho^0)$ {\em uniformly}. This is
    strongly dominated by $\rho^+$ over $\rho^-$ for DY and VBF and is a
    consequence of the proton PDFs.
\end{itemize}

The DY cross sections vary roughly as $1/g^2_{\rho_H}$ for $M_{\rho_H}$ fixed
near $2\,\tev$. On the other hand, the VBF rate for $\rho_H \to VV$ varies as
$g^2_{\rho_H}$ for fixed $M_{\rho_H}$. Then, e.g., $g_{\rho_H} = 2.73$ gives
a 75\% larger production rate for $\rho_H \to VV$ and a width half as large.

\begin{table}[!t]
     \begin{center}{
  \begin{tabular}{|c|c|c|c|c|c|}
  \hline
$\sqrt{s}$ & $M_{\rho_H}$ (GeV)& $\sigma(\rho_H^\pm)_{DY+VBF}$ (fb) & 
$\sigma(\rho_H^0)_{DY+VBF}$  (fb)& $\sigma(a_H^\pm)$ (fb) & $\sigma(a_H^0)$ (fb)\\
  \hline\hline
8 & 1800 & 1.53 $+$ 0.36 & 0.74 $+$ 0.18  & 0.71 & 0.37  \\ 
8 & 1900 & 1.05 $+$ 0.24 & 0.50 $+$ 0.12  & 0.51 & 0.27  \\
8 & 2000 & 0.73 $+$ 0.15 & 0.36 $+$ 0.075 & 0.36 & 0.17  \\
\hline\hline
13 & 1800 & 7.61 $+$ 3.67 & 3.74 $+$ 1.93  & 4.65 & 2.23  \\ 
13 & 1900 & 5.74 $+$ 2.62 & 2.81 $+$ 1.37  & 3.16 & 1.69  \\
13 & 2000 & 4.37 $+$ 1.90 & 2.16 $+$ 0.99  & 2.39 & 1.27  \\
\hline
 \end{tabular}}
 \caption{Production cross sections at the LHC of the isovector bosons
   $\rho_H$ and $a_H$ for $g_{\rho_H} = 3.862$ and $M_{a_H} = 1.05
   M_{\rho_H}$ ($\rho_H^\pm = \rho_H^+ + \rho_H^-)$. The individual DY $+$
   VBF contributions are given for $\rho_H$; the VBF rates for $a_H$ are very
   small and not given. As explained in the text, $g_{\rho_H} = 2.73$ gives
   75\% larger cross sections and widths half as large for $\rho_H \to VV$.
   No $K$-factor has been applied.
   \label{tab:Xsections}}
 \end{center}
 \end{table}

\section*{4. Comments and Predictions}

In this paper we proposed that the excess diboson events near $M_{VV} =
2\,\tev$ reported by ATLAS and CMS are due to production of isovector bosons,
$\rho_H$ and $a_H$, associated with new strong dynamics that make the Higgs
boson a light composite state. We focused on two types of models that have a
custodial $SU(2)$-isospin symmetry and approximate left-right symmetry. We
believe our results are equally applicable to both types. Here we make some
comments and predictions implied by them and which can be tested in the next
couple of years.

\begin{itemize}
  
\item[1)]The $\rho_H^0,a_H^0 \to ZZ$ decays are isospin-violating and their
  rates are very small. Therefore, the $ZZ$ signals claimed by ATLAS and CMS
  will be understood to have one or two misidentified $Z$-bosons. (A
  possibility we have not considered is the production of an $I=0$ scalar,
  $f_0$-like, which could decay to $ZZ$. Its production would have to be via
  VBF.)
  
\item[2)] It is difficult for us to explain cross sections greater than
  a few~fb for individual diboson ($WW$ or $WZ$) production at
  $\sqrt{s}=8\,\tev$. Therefore, we expect that, should these signals be
  confirmed in Run~2, they will be seen to have been up-fluctuations in
  Run~1, something quite familiar in the history of particle physics,
  including the discovery of the Higgs
  boson~\cite{Aad:2012tfa,Chatrchyan:2012ufa}.
  
\item[3)] There must be semileptonic $VV$ events, their present spotty
  evidence being a consequence of low statistics. The $\ell\nu \bar q q$
  events should have $\sigma(\ellp)/\sigma(\ellm) \simeq 2$.
  
\item[4)] The $\rho_H$ width is almost entirely due to strong-interaction
  decays to $VV$ and is $\sim 200\,\gev$ with our parameters. Presumably, it
  would be best measured in semileptonic $VV$ events.
  
\item[5)] $\rho_H \to VV$ decays involve a pair of longitudinally-polarized
  weak bosons. Note that boosted $V_L$ tend to produce quark-subjets that
  have more equal momenta along the parent $V$-direction than do boosted
  $V_T$. Also see Ref.~\cite{Cui:2010km}.

\item[6)] A measurement of the $\rho_H$ width {\em {\underbar{is}}} a
  measurement of $VV$ polarizations. A large width can be due only to strong
  dynamics, hence emission of $V_L V_L$. A small width is an electroweak decay
  involving $V_L V_T$ or $V_T V_T$. 
  
\item[7)] The $VH$ signal should strengthen with more data. It is entirely
  due to the strong decay $a_H \to VH$, hence it involves $V_L$ and a large
  width. In our model $\Gamma(a_H) \cong \Gamma(\rho_H)$.

\item[8)] There should be forward jets from VBF in $\rho_H \to VV$, but not
  in $a_H \to VH$.

\item[9)] Finally, if $H$ is a PGB, there likely are top and $W$-partners
  that keep it light. They are not hadrons of the new strong dynamics and,
  so, are surely lighter than $\rho_H, a_H$. They should show up soon. There
  are {\em no} top and $W$-partners needed in the strong-ETC model and there
  aren't any.

\end{itemize}

\section*{Acknowledgments}

We have benefited from conversations with Tom Appelquist, Gustaaf Brooijmans,
John Butler, Louis Helary, Adam Martin, Chris Pollard, Bing Zhou and Junjie
Zhu. KL gratefully acknowledges support of this project by a CERN Scientific
Associateship and by the Labex ENIGMASS.  He thanks the CERN Theory Group and
Laboratoire d'Annecy-le-Vieux de Physique Th\'eorique (LAPTh) for their
gracious hospitality. Our research is supported in part by the
U.S.~Department of Energy under Grant No.~DE-SC0010106.


\bibliography{Composite-Higgs-Resonances}

\providecommand{\href}[2]{#2}\begingroup\raggedright\begin{thebibliography}{10}

\bibitem{Aad:2015owa}
{\bf ATLAS} Collaboration, G.~Aad {\em et.~al.}, ``{Search for high-mass
  diboson resonances with boson-tagged jets in proton-proton collisions at
  $\sqrt{s}$ = 8 TeV with the ATLAS detector},''
  \href{http://xxx.lanl.gov/abs/1506.00962}{ 1506.00962}.

\bibitem{Khachatryan:2014gha}
{\bf CMS} Collaboration, V.~Khachatryan {\em et.~al.}, ``{Search for massive
  resonances decaying into pairs of boosted bosons in semi-leptonic final
  states at $\sqrt{s} =$ 8 TeV},'' {\em JHEP} {\bf 08} (2014) 174,
  \href{http://xxx.lanl.gov/abs/1405.3447}{ 1405.3447}.

\bibitem{Khachatryan:2014hpa}
{\bf CMS} Collaboration, V.~Khachatryan {\em et.~al.}, ``{Search for massive
  resonances in dijet systems containing jets tagged as W or Z boson decays in
  pp collisions at $ \sqrt{s} $ = 8 TeV},'' {\em JHEP} {\bf 08} (2014) 173,
  \href{http://xxx.lanl.gov/abs/1405.1994}{ 1405.1994}.

\bibitem{Aad:2014xka}
{\bf ATLAS} Collaboration, G.~Aad {\em et.~al.}, ``{Search for resonant diboson
  production in the $\mathrm {\ell \ell }q\bar{q}$ final state in $pp$
  collisions at $\sqrt{s} = 8$ TeV with the ATLAS detector},'' {\em Eur. Phys.
  J.} {\bf C75} (2015), no.~2, 69, \href{http://xxx.lanl.gov/abs/1409.6190}{
  1409.6190}.

\bibitem{Aad:2015ufa}
{\bf ATLAS} Collaboration, G.~Aad {\em et.~al.}, ``{Search for production of
  $WW/WZ$ resonances decaying to a lepton, neutrino and jets in $pp$ collisions
  at $\sqrt{s}=8$ TeV with the ATLAS detector},'' {\em Eur. Phys. J.} {\bf C75}
  (2015), no.~5, 209, \href{http://xxx.lanl.gov/abs/1503.04677}{ 1503.04677}.

\bibitem{CMSPAS:2014yyy}
{\bf CMS} Collaboration, ``Search for massive WH resonances decaying to the
  $\ell\nu \bar bb$ final state in the boosted regime at $\sqrt{s} = 8\,\tev$,
  CMS PAS EXO-14-010.''

\bibitem{Aad:2015yza}
{\bf ATLAS} Collaboration, G.~Aad {\em et.~al.}, ``{Search for a new resonance
  decaying to a $W$ or $Z$ boson and a Higgs boson in the $\ell \ell/ \ell \nu/
  \nu \nu + b \bar{b}$ final states with the ATLAS Detector},'' {\em Eur. Phys.
  J.} {\bf C75} (2015), no.~6, 263, \href{http://xxx.lanl.gov/abs/1503.08089}{
  1503.08089}.

\bibitem{Cheung:2015nha}
K.~Cheung, W.-Y. Keung, P.-Y. Tseng, and T.-C. Yuan, ``{Interpretations of the
  ATLAS Diboson Anomaly},'' \href{http://xxx.lanl.gov/abs/1506.06064}{
  1506.06064}.

\bibitem{Dobrescu:2015qna}
B.~A. Dobrescu and Z.~Liu, ``{A W' boson near 2 TeV: predictions for Run 2 of
  the LHC},'' \href{http://xxx.lanl.gov/abs/1506.06736}{ 1506.06736}.

\bibitem{Abe:2015uaa}
T.~Abe, T.~Kitahara, and M.~M. Nojiri, ``{Prospects for Spin-1 Resonance Search
  at 13 TeV LHC and the ATLAS Diboson Excess},''
  \href{http://xxx.lanl.gov/abs/1507.01681}{ 1507.01681}.

\bibitem{Brehmer:2015cia}
J.~Brehmer, J.~Hewett, J.~Kopp, T.~Rizzo, and J.~Tattersall, ``{Symmetry
  Restored in Dibosons at the LHC?},''
  \href{http://xxx.lanl.gov/abs/1507.00013}{ 1507.00013}.

\bibitem{Thamm:2015csa}
A.~Thamm, R.~Torre, and A.~Wulzer, ``{A composite Heavy Vector Triplet in the
  ATLAS di-boson excess},'' \href{http://xxx.lanl.gov/abs/1506.08688}{
  1506.08688}.

\bibitem{Franzosi:2015zra}
D.~B. Franzosi, M.~T. Frandsen, and F.~Sannino, ``{Diboson Signals via Fermi
  Scale Spin-One States},'' \href{http://xxx.lanl.gov/abs/1506.04392}{
  1506.04392}.

\bibitem{Carmona:2015xaa}
A.~Carmona, A.~Delgado, M.~Quiros, and J.~Santiago, ``{Diboson resonant
  production in non-custodial composite Higgs models},''
  \href{http://xxx.lanl.gov/abs/1507.01914}{ 1507.01914}.

\bibitem{Cacciapaglia:2015nga}
G.~Cacciapaglia, A.~Deandrea, and M.~Hashimoto, ``{A scalar hint from the
  diboson excess?},'' \href{http://xxx.lanl.gov/abs/1507.03098}{ 1507.03098}.

\bibitem{Sanz:2015zha}
V.~Sanz, ``{On the compatibility of the diboson excess with a gg-initiated
  composite sector},'' \href{http://xxx.lanl.gov/abs/1507.03553}{ 1507.03553}.

\bibitem{Contino:2011np}
R.~Contino, D.~Marzocca, D.~Pappadopulo, and R.~Rattazzi, ``{On the effect of
  resonances in composite Higgs phenomenology},'' {\em JHEP} {\bf 10} (2011)
  081, \href{http://xxx.lanl.gov/abs/1109.1570}{ 1109.1570}.

\bibitem{Marzocca:2012zn}
D.~Marzocca, M.~Serone, and J.~Shu, ``{General Composite Higgs Models},'' {\em
  JHEP} {\bf 08} (2012) 013, \href{http://xxx.lanl.gov/abs/1205.0770}{
  1205.0770}.

\bibitem{Bellazzini:2012tv}
B.~Bellazzini, C.~Csaki, J.~Hubisz, J.~Serra, and J.~Terning, ``{Composite
  Higgs Sketch},'' {\em JHEP} {\bf 11} (2012) 003,
  \href{http://xxx.lanl.gov/abs/1205.4032}{ 1205.4032}.

\bibitem{Lane:2014vca}
K.~Lane, ``{A composite Higgs model with minimal fine-tuning: The large-$N$ and
  weak-technicolor limit},'' {\em Phys.Rev.} {\bf D90} (2014), no.~9, 095025,
  \href{http://xxx.lanl.gov/abs/1407.2270}{ 1407.2270}.

\bibitem{Bando:1987br}
M.~Bando, T.~Kugo, and K.~Yamawaki, ``{Nonlinear Realization and Hidden Local
  Symmetries},'' {\em Phys. Rept.} {\bf 164} (1988) 217--314.

\bibitem{Bellazzini:2014yua}
B.~Bellazzini, C.~Csáki, and J.~Serra, ``{Composite Higgses},'' {\em
  Eur.Phys.J.} {\bf C74} (2014), no.~5, 2766,
  \href{http://xxx.lanl.gov/abs/1401.2457}{ 1401.2457}.

\bibitem{Panico:2015jxa}
G.~Panico and A.~Wulzer, ``{The Composite Nambu-Goldstone Higgs},''
  \href{http://xxx.lanl.gov/abs/1506.01961}{ 1506.01961}.

\bibitem{Nambu:1961tp}
Y.~Nambu and G.~Jona-Lasinio, ``{Dynamical Model of Elementary Particles Based
  on an Analogy with Superconductivity. 1.},'' {\em Phys.Rev.} {\bf 122} (1961)
  345--358.

\bibitem{Nambu:1961fr}
Y.~Nambu and G.~Jona-Lasinio, ``{DYNAMICAL MODEL OF ELEMENTARY PARTICLES BASED
  ON AN ANALOGY WITH SUPERCONDUCTIVITY. II},'' {\em Phys.Rev.} {\bf 124} (1961)
  246--254.

\bibitem{Dimopoulos:1979sp}
S.~Dimopoulos, ``{Technicolored Signatures},'' {\em Nucl. Phys.} {\bf B168}
  (1980) 69--92.

\bibitem{Dimopoulos:1980yf}
S.~Dimopoulos, S.~Raby, and G.~L. Kane, ``{Experimental Predictions from
  Technicolor Theories},'' {\em Nucl. Phys.} {\bf B182} (1981) 77--103.

\bibitem{Kennedy:1988sn}
D.~C. Kennedy and B.~W. Lynn, ``{Electroweak Radiative Corrections with an
  Effective Lagrangian: Four Fermion Processes},'' {\em Nucl. Phys.} {\bf B322}
  (1989) 1.

\bibitem{Peskin:1990zt}
M.~E. Peskin and T.~Takeuchi, ``A new constraint on a strongly interacting
  Higgs sector,'' {\em Phys. Rev. Lett.} {\bf 65} (1990) 964--967.

\bibitem{Golden:1990ig}
M.~Golden and L.~Randall, ``Radiative corrections to electroweak parameters in
  technicolor theories,'' {\em Nucl. Phys.} {\bf B361} (1991) 3--23.

\bibitem{Holdom:1990tc}
B.~Holdom and J.~Terning, ``Large corrections to electroweak parameters in
  technicolor theories,'' {\em Phys. Lett.} {\bf B247} (1990) 88--92.

\bibitem{Altarelli:1991fk}
G.~Altarelli, R.~Barbieri, and S.~Jadach, ``Toward a model independent analysis
  of electroweak data,'' {\em Nucl. Phys.} {\bf B369} (1992) 3--32.

\bibitem{Casalbuoni:1995qt}
R.~Casalbuoni {\em et.~al.}, ``{Degenerate BESS Model: The possibility of a low
  energy strong electroweak sector},'' {\em Phys. Rev.} {\bf D53} (1996)
  5201--5221, \href{http://xxx.lanl.gov/abs/hep-ph/9510431}{ hep-ph/9510431}.

\bibitem{Lane:2009ct}
K.~Lane and A.~Martin, ``{An Effective Lagrangian for Low-Scale Technicolor},''
  {\em Phys. Rev.} {\bf D80} (2009) 115001,
  \href{http://xxx.lanl.gov/abs/0907.3737}{ 0907.3737}.

\bibitem{Lane:2002sm}
K.~Lane and S.~Mrenna, ``{The Collider phenomenology of technihadrons in the
  technicolor straw man model},'' {\em Phys.Rev.} {\bf D67} (2003) 115011,
  \href{http://xxx.lanl.gov/abs/hep-ph/0210299}{ hep-ph/0210299}.

\bibitem{Sjostrand:2006za}
T.~Sjostrand, S.~Mrenna, and P.~Skands, ``PYTHIA 6.4 physics and manual,'' {\em
  JHEP} {\bf 05} (2006) 026, \href{http://xxx.lanl.gov/abs/hep-ph/0603175}{
  hep-ph/0603175}.

\bibitem{Aguilar-Saavedra:2015rna}
J.~A. Aguilar-Saavedra, ``{Triboson interpretations of the ATLAS diboson
  excess},'' \href{http://xxx.lanl.gov/abs/1506.06739}{ 1506.06739}.

\bibitem{Aad:2012tfa}
{\bf ATLAS} Collaboration, G.~Aad {\em et.~al.}, ``{Observation of a new
  particle in the search for the Standard Model Higgs boson with the ATLAS
  detector at the LHC},'' {\em Phys.Lett.} {\bf B716} (2012) 1--29,
  \href{http://xxx.lanl.gov/abs/1207.7214}{ 1207.7214}.

\bibitem{Chatrchyan:2012ufa}
{\bf CMS} Collaboration, S.~Chatrchyan {\em et.~al.}, ``{Observation of a new
  boson at a mass of 125 GeV with the CMS experiment at the LHC},'' {\em
  Phys.Lett.} {\bf B716} (2012) 30--61,
  \href{http://xxx.lanl.gov/abs/1207.7235}{ 1207.7235}.

\bibitem{Cui:2010km}
Y.~Cui, Z.~Han, and M.~D. Schwartz, ``{W-jet Tagging: Optimizing the
  Identification of Boosted Hadronically-Decaying W Bosons},'' {\em Phys. Rev.}
  {\bf D83} (2011) 074023, \href{http://xxx.lanl.gov/abs/1012.2077}{
  1012.2077}.

\end{thebibliography}\endgroup
\bibliographystyle{utcaps}
\end{document}